\documentclass[twocolumn,a4paper,superscriptaddress,floatfix,showpacs]{revtex4}
\usepackage{amsmath,amssymb,epsfig,color,textcase}

\begin{document}

\title{Magnetic moment suppression in Ba$_3$CoRu$_2$O$_9$: hybridization effect}
\author{S.V.~Streltsov}
\affiliation{Institute of Metal Physics, S.Kovalevskoy St. 18, 620990 Ekaterinburg, Russia}
\email{streltsov@imp.uran.ru}


\date{\today}

\begin{abstract}
An unusual orbital state was recently proposed to explain the magnetic and 
transport properties of Ba$_3$CoRu$_2$O$_9$ [Phys. Rev. B. {\bf 85}, 041201 (2012)].
We show that this state contradicts to the first Hund's rule and
does not realize in the system under consideration because of a too small
crystal-field splitting in the $t_{2g}$ shell. A strong suppression of the
local magnetic moment in Ba$_3$CoRu$_2$O$_9$ is attributed to a strong hybridization
between the Ru 4$d$ and O 2$p$ states.
\end{abstract}

\maketitle

\section{Introduction \label{intro}}
The 4$d$ and 5$d$ based transition metal compounds are widely investigated in the 
last few years. A larger spatial extension of the 4$d$ and 5$d$ wave 
functions and a substantial spin-orbit coupling makes 
them quite different from the 3$d$ analogues in terms of the electronic and especially
magnetic properties. 

This for instance results in unusual zigzag antiferromagnetic order and
a quasimolecular orbital state in Na$_2$IrO$_3$,~\cite{Mazin2012}
unconventional magnetic properties and charge-ordered state 
sensitive to irradiation in Ba$_3$NaRu$_2$O$_9$,~\cite{Kimber2012} formation of the
spin singlets in La$_4$Ru$_2$O$_{10}$,~\cite{Wu2006} or suppression of the magnetic 
moments in such compounds as Ba$_4$Ru$_3$O$_{10}$~\cite{Streltsov2012a} 
and Ba$_2$NaOsO$_6$.~\cite{Erickson2007}
A strong reduction of the local magnetic moment was also found in 
Ba$_3$CoRu$_2$O$_9$~\cite{Lightfoot1990,Rijssenbeek1999} and recently attributed 
to a special type of the orbital order, which leads to an unusual
orbital filling.~\cite{Zhou2012}

Ba$_3$CoRu$_2$O$_9$ is a semiconductor~\cite{Rijssenbeek1998} 
and experiences a magnetic transition at $T_N=93$~K, which is accompanied 
by the changes in the crystal symmetry from orthorhombic (Cmcm) in the low temperature (LT)
phase to hexagonal (P6$_3$/mmc) at higher temperatures.~\cite{Zhou2012}
The Ru$^{5+}$ ions are in the $d^3$ electronic configurations, while 
the Co ions show 2+ oxidation state with seven 3$d$ electrons. 
Since the RuO$_6$ octahedra are strongly distorted~\cite{Lightfoot1990} one may expect 
that the orbital moment is quenched and the total magnetic moment
is defined by the spin component only. However, the neutron measurements 
show that the local magnetic moment on the Ru in the LT phase is 
1.17-1.45 $\mu_B$~according to Ref.~\onlinecite{Lightfoot1990,Rijssenbeek1999}, much
smaller than 3 $\mu_B$, expected for S=3/2 from the naive atomic
consideration. In contrast the local magnetic moment on Co was found to be 
2.71-2.75 $\mu_B$~\cite{Lightfoot1990,Rijssenbeek1998}, which
is close to the spin only value 3 $\mu_B$ for Co$^{2+}$ (S=3/2).

The reduction of the local magnetic moment on Ru$^{5+}$ was explained in 
Ref.~\onlinecite{Zhou2012} as a result of the stabilization of an unconventional orbital
state, when one of the $t_{2g}$ orbitals is completely filled (with spin up and
down electrons), so that the total spin is S=1/2 per Ru site (due to the remaining
unpaired electron). This, however, contradicts to the first Hund's rule, which states
that the term with maximum spin (i.e. S=3/2 in the case of $t^3_{2g}$ 
configuration) has the lowest total energy.
In a simple ionic model this ``anti-Hund's rule''state is possible if the crystal 
field splitting in the $t_{2g}$ shell is larger than 2J$_H$, which is
quite unlikely since J$_H$ is $\sim$ 0.7 eV for the Ru.~\cite{Lee2006}

In the present paper we investigate the electronic and magnetic structure 
of Ba$_3$CoRu$_2$O$_9$ with the use of the band structure calculations and 
show that the splitting in the $t_{2g}$ shell does not
exceed 108 meV. As a result the unconventional orbital state proposed
in Ref.~\onlinecite{Zhou2012} is not realized. The suppression of the value of the local 
magnetic moment on Ru is explained by the hybridization effects with the O $2p$ states.
Substantial hybridization between the Ru $4d$ and O $2p$ states leads to a localization of
the electrons not on the atomic, but on the Wannier orbitals, with a large contribution 
coming from the non magnetic O $2p$ states.

\begin{figure}[b!]
 \centering
 \includegraphics[clip=false,width=0.4\textwidth]{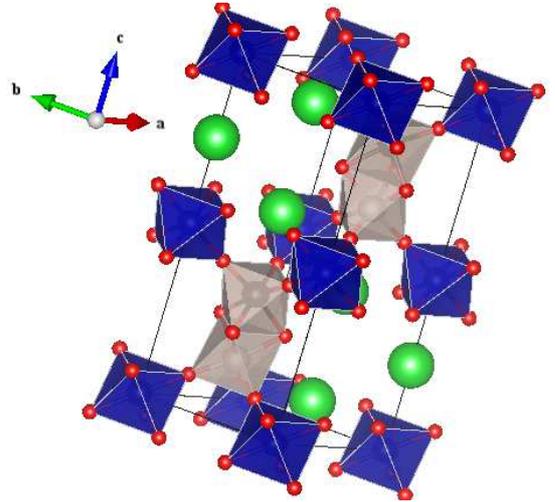}
\caption{\label{cryst.str}(color online). The crystal structure of
the Ba$_3$CoRu$_2$O$_9$. Oxygen ions are shown in red, Ba - in 
green. Co$^{2+}$ and Ru$^{5+}$ ions are placed inside of 
the oxygen octahedra painted in blue and grey respectively. 
The image was generated using VESTA software.~\cite{MommaK.Izumi2011}}
\end{figure}

\section{Calculation and crystal structure details}
The linearized muffin-tin orbitals method (LMTO) was used in the 
calculations~\cite{Andersen1984} with the von Barth-Hedin
version of the exchange correlation potential.~\cite{Barth1972}
We investigated the effect of the possible strong Coulomb interaction
on the $d$ shells of the Ru and Co ions with the mean-field
LSDA+U method.~\cite{Anisimov1997} The on-site Coulomb 
repulsion parameter U and the intra-atomic Hund's rule exchange J$_H$
were chosen as following:
U(Co)=6 eV, J$_H$(Co)=1 eV,~\cite{Dyachenko2012} U(Ru)=3 eV, 
J$_H$(Ru)=0.7 eV~\cite{Lee2006}. In order to check the stability of the results
these parameters were varied as it will be discussed in what follows.
We used the mesh of the 144 k-points in the full Brillouin zone
in the course of the calculations.

The inter-site exchange interaction parameters were calculated for the
Heisenberg model written as
\begin{equation}
\label{Heisenberg}
H = \sum_{ij} J \vec S_i \vec S_j,
\end{equation}
(i.e. each site is counted twice in the summation) using the Green's function method 
described elsewhere.~\cite{LEIP1}

The crystal structure was taken from Ref.~\onlinecite{Lightfoot1990} for T=2 K
and is shown in Fig.~\ref{cryst.str}. The Ru ions are placed in the center of the RuO$_6$ octahedra, 
which form dimers sharing their faces. These dimers are directed along the $c$ axis,
but the Ru--Ru dimer and the Ba--Ba pairs are alternating in the $c$ direction.
Three neighboring dimers lying in the same $ab$ plane are interconnected by the 
CoO$_6$ octahedron. The CoO$_6$ and RuO$_6$ octahedra share one of the corners.

\section{\label{LDA.sec}LDA results}
We start with the conventional non-magnetic calculations performed in the 
Local density approximation (LDA). Using the Wannier function projection 
technique~\cite{Streltsov2005} one may obtain the values of the crystal-field 
splitting in the Ru $t_{2g}$ sub-shell. Diagonalizing a small on-site $t_{2g}$--$t_{2g}$
Hamiltonian we found that the degeneracy of the Ru $t_{2g}$ states is 
lifted due to a low symmetry (four out of six Ru-O bond lengths are different). 
The crystal-field splitting is 58 meV (between the lowest in energy and middle
states) and 108 meV (between the middle and highest in energy orbitals). 
This is much smaller than $2J_H \approx$1.4 eV needed to stabilize the 
``anti-Hund's rule'' state for the $d^3$ configuration with the total 
spin moment S=1/2 per site.

However, a close inspection of the projected Hamiltonian show
that there are other terms, even larger than the on-site splitting
in the Ru $t_{2g}$ shell. These are the hoppings between the Ru $t_{2g}$ orbitals 
centered on different sites (exceed 290 meV), and off-diagonal matrix
elements between the Ru $t_{2g}$ and Co $e_g$ states ($\sim$100 meV). Thus,
one may expect that the band structure in the vicinity of the
Fermi level is rather governed by the inter-site, not on-site
elements of the Hamiltonian, which is obviously a consequence
of the dimerized crystal structure.

\begin{figure}[t!]
 \centering
 \includegraphics[clip=false,angle=270,width=0.4\textwidth]{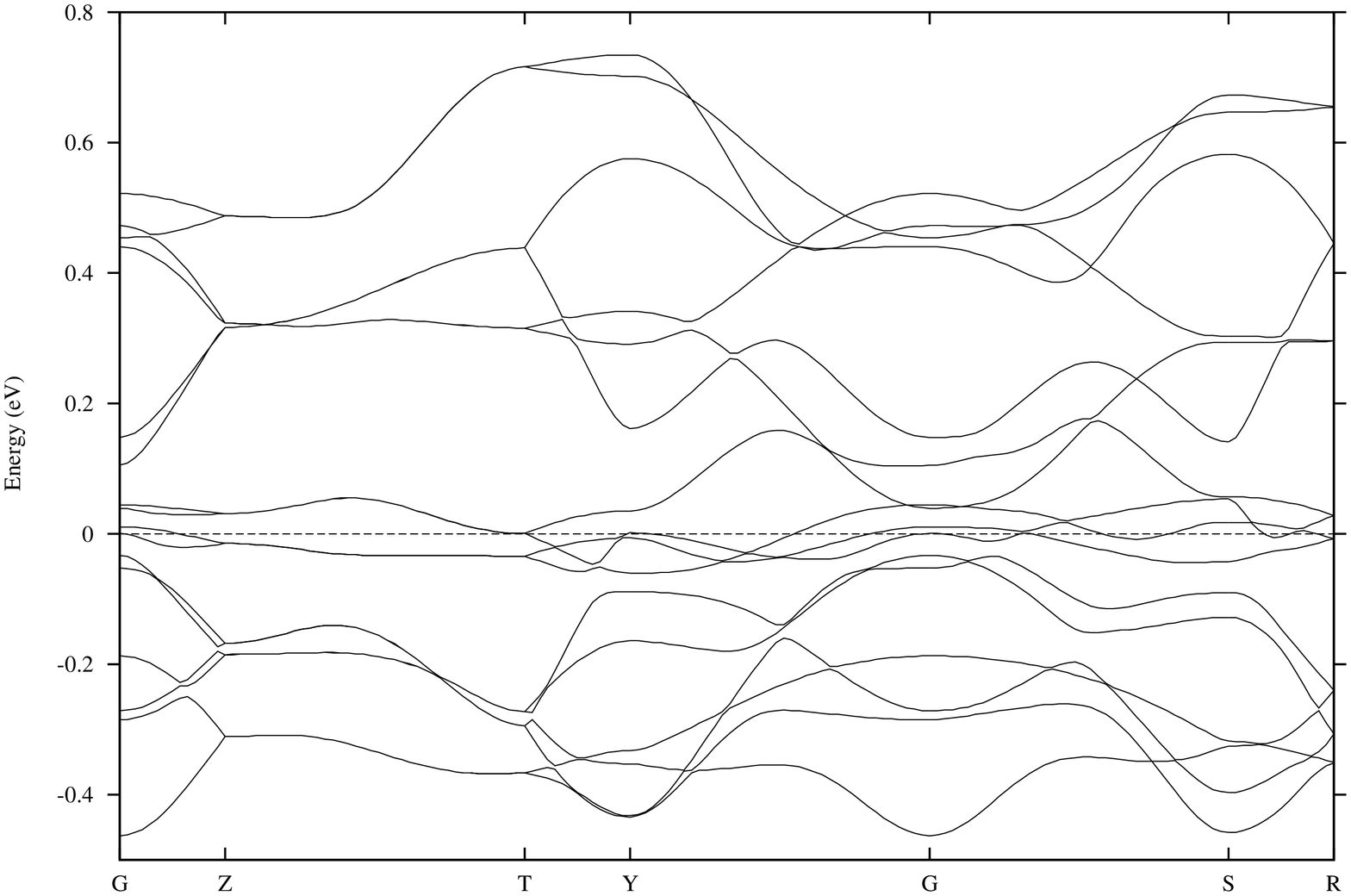}
 \includegraphics[clip=false,angle=270,width=0.4\textwidth]{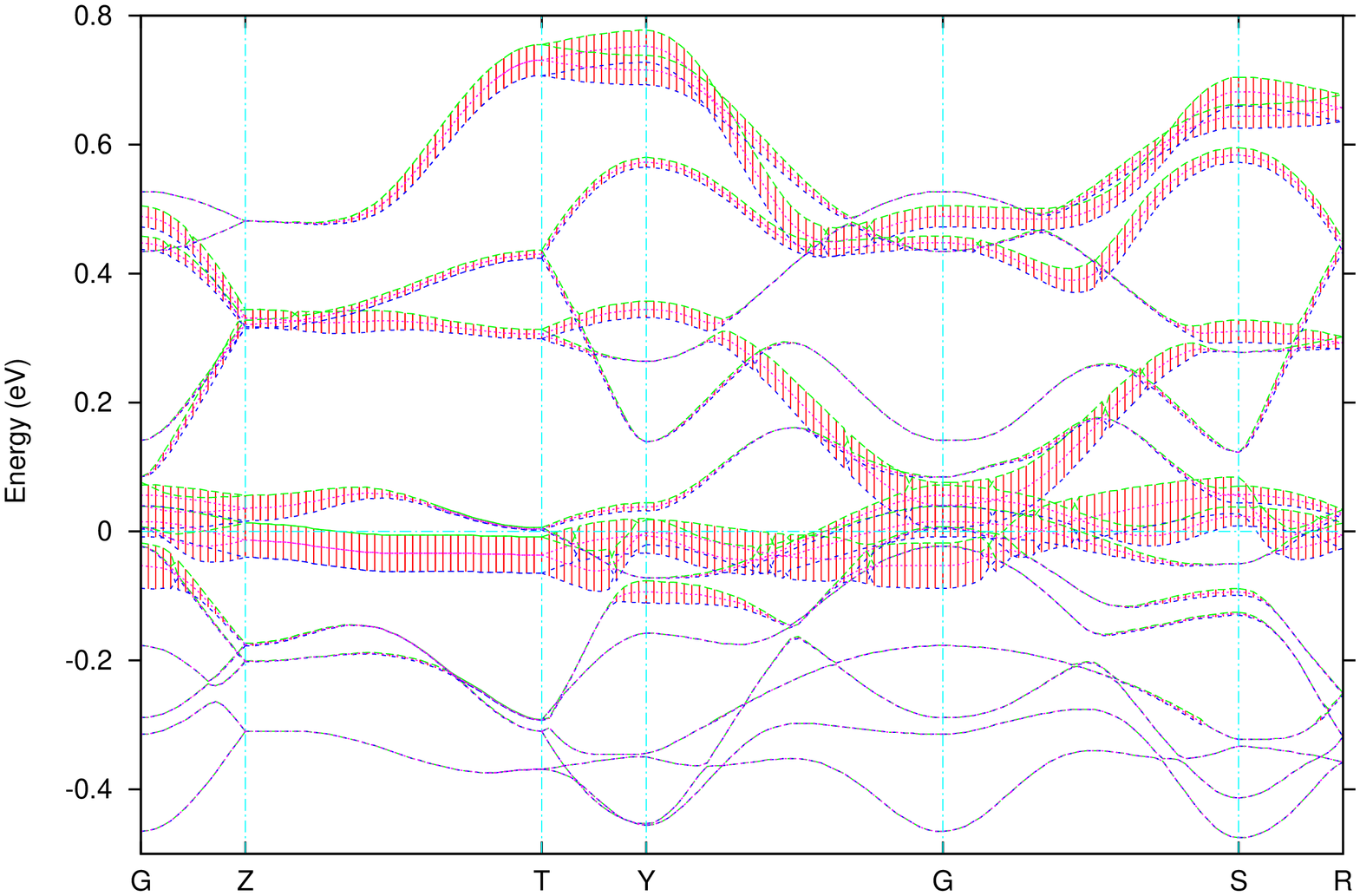}
\caption{\label{LDA-bands} The band structure for 
Ba$_3$CoRu$_2$O$_9$, obtained in the LDA calculation. In the lower panel 
the contribution coming from the Co $e_g$ states is shown (so called 
``fatbands''): the broader is a given band in a certain $k-$point the larger
is a contribution from the Co $e_g$ states. The local coordinate system
with axis pointing to oxygens was used to identify the Co $e_g$ states. 
The Fermi level is set to zero.}
\end{figure}

The LDA band structure is shown in Fig.~\ref{LDA-bands}. One may
see that there are essentially three branches of the bands.
Four bands placed exactly on the Fermi level mostly 
have the Co $e_g$ character (see lower panel in Fig.~\ref{LDA-bands}). 
Each band is two times degenerate in the ZT direction, due to the fact 
that there are two formula units in the unit cell. The Co $e_g$ 
bands are flat, which is related to the feature of the crystal
structure: the CoO$_6$ octahedra are not directly connected with 
each other, only via RuO$_6$. Moreover, the Co-Ru-Co angle is close
to 90$^{\circ}$. The flat Co $e_g$ bands provide enormous density
of states (DOS) at the Fermi level $\sim$ 40 states/(eV*f.u.),
which results as we will see below in the magnetic instability
according to the Stoner criteria.

The lowermost six bands, lying below the Co $e_g$ bands 
correspond to the Ru $t_{2g}$ bonding states in the Ru-Ru dimer. 
The lowest, at $\sim$-0.3 eV in the ZT direction, are the $a_{1g}$ orbitals, 
and the rest have the $e_g^{\pi}$ symmetry. Two $a_{1g}$ orbitals of the
neighboring Ru ions in the shared faces geometry directed exactly to 
each other. This leads to a large hopping between those wave functions
 and as a result the bonding-antibonding splitting for the $a_{1g}$ 
orbitals is much larger than for the $e_g^{\pi}$ ones.

The uppermost six bands (spread from $\sim$ 0.08 eV to 0.8 eV) are 
the Ru $t_{2g}$ antibonding bands with an admixture of the 
Co $e_g$ states.

\section{LSDA results}
The large DOS at the Fermi level in the nonmagnetic LDA calculation 
leads to the magnetic instability. We used the Local spin density 
approximation (LSDA) as a simplest method to study magnetic properties of 
Ba$_3$CoRu$_2$O$_9$. This approach was shown to provide an adequate description of 
the Ru-based compounds.~\cite{Etz2012,Streltsov2012a}
Experimentally determined magnetic structure~\cite{Lightfoot1990} was 
used in the present calculations. It was shown that the Ru ions forming 
first magnetic lattice are paired antiferromagnetically in the 
dimers. The interdimer coupling in the $c$ direction is also 
antiferromagnetic. The second magnetic lattice consists
of the Co ions, which are antiferromagnetically paired
in the $c$ and $b$ directions, but ferromagnetically
along the $a$ direction (see Fig. 4 in Ref.~\onlinecite{Lightfoot1990}).

\begin{figure}[t!]
 \centering
 \includegraphics[clip=false,width=0.4\textwidth]{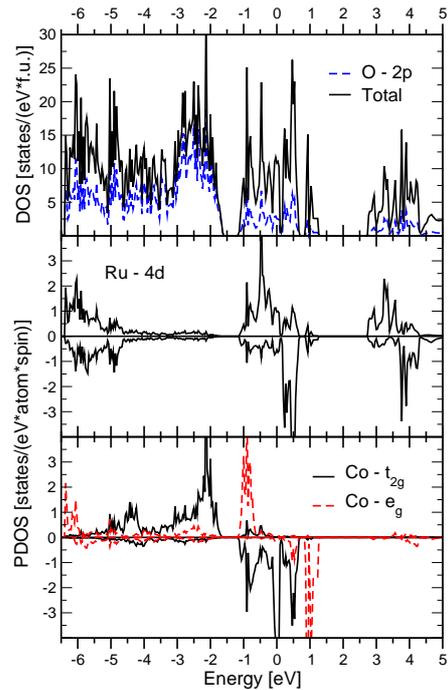}
\caption{\label{LSDA} (color online).  
The total (TDOS) and partial (PDOS) density
of states for Ba$_3$CoRu$_2$O$_9$, obtained in the LSDA calculation. 
The positive (negative) values on the lowest two plots
correspond to spin majority (minority). The local coordinate system, 
when axes are directed to oxygens, was chosen to identify
the Co $t_{2g}$ and e$_g$ states. The Fermi level is set to zero.}
\end{figure}

The results of the LSDA calculation are presented in Fig.~\ref{LSDA}.
The magnetic interaction splits the Co $3d$ states with different
spins on $\approx$2 eV, which results in the formation of a sizable 
spin moment on the Co ion (2.55~$\mu_B$). However, shifting the Co 
$e_g$ states away from the Fermi level the magnetic splitting
puts the Co $t_{2g}\downarrow$ states on their place. As a result
in the LSDA Ba$_3$CoRu$_2$O$_9$ stays metallic in contrast to the
experimental observations.~\cite{Zhou2012} The magnetic
moment on the Ru 1.1$\mu_B$ is close to the experimentally measured
value.~\cite{Rijssenbeek1999,Lightfoot1990} In order to stabilize an 
insulating ground state in the following we will apply
the LSDA+U method, which allows to take into account
strong on-site Coulomb correlations in a  mean field way.~\cite{Anisimov1997} 

\section{LSDA+U results}

Since there are two transition metal ions in Ba$_3$CoRu$_2$O$_9$,
for which an account of the strong Coulomb correlations can be
important, we applied the $U-$correction step by step. 
\begin{figure}[t!]
 \centering
 \includegraphics[clip=false,width=0.4\textwidth]{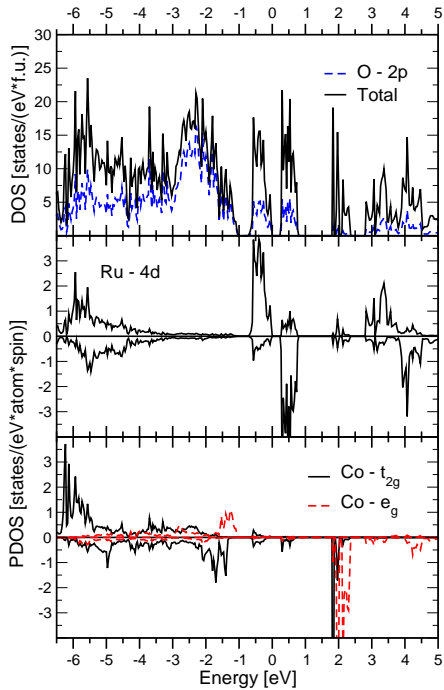}
\caption{\label{LSDAU} (color online).  
The total (TDOS) and partial (PDOS) density
of states for Ba$_3$CoRu$_2$O$_9$, obtained in the LSDA+U$_{Co}$ 
calculation.  The positive (negative) values on the lowest two plots
correspond to spin majority (minority). The local coordinate system, 
when axes are directed to oxygens, was chosen to identify
the Co $t_{2g}$ and e$_g$ states. The Fermi level is set to zero.}
\end{figure}

First of all U = 6 eV and J$_H=1$ eV were applied for the Co $3d$ shell only
 (will be cited as LSDA+U$_{Co}$ in what follows). This results in the magnetic moments
$|m_{Ru}| =1.47$~$\mu_B$, $|m_{Co}| =2.61$~$\mu_B$ and 
the band gap 0.25 eV. This agrees both with experimental estimations
of the magnetic moment on Ru of 1.17-1.45 $\mu_B$~\cite{Rijssenbeek1999,Lightfoot1990}
and semiconducting resistivity temperature 
dependence.~\cite{Zhou2012} It is important to note that the spin density
is almost homogeneously distributed over all $t_{2g}$ orbitals of the 
Ru$^{5+}$ ion leading to the orbital polarizations $|m_{t_{2g,1}}|=$0.49~$\mu_B$,
$|m_{t_{2g,2}}|=$0.47~$\mu_B$, and $|m_{t_{2g,3}}|=$0.43~$\mu_B$ (the rest 
0.08~$\mu_B$ comes from the $e_g$ orbitals). This is obviously due to the fact,
that the crystal field splitting in the $t_{2g}$ shell significantly smaller
than the intra-atomic Hund's rule exchange coupling $J_H$, as it was shown in 
Sec.~\ref{LDA.sec}.

 The results obtained are stable with respect
to the small variation of the U and J$_H$ parameters. The decrease of the 
J$_H$ on 20$\%$ does not change neither the spin moments nor the 
band gap value. The calculation with U=5 eV decreases
the band gap on 0.01 eV and the spin moments on 0.01 $\mu_B$
and 0.04 $\mu_B$ for the Ru and Co atoms respectively. 
Thus, the main effect of the $U-$correction on the Co $3d$ states 
is to push them away from the Fermi level and stabilize
the insulating ground state. This can be seen in Fig.~\ref{LSDAU}.

\begin{figure}[t!]
 \centering
 \includegraphics[clip=false,angle=270,width=0.5\textwidth]{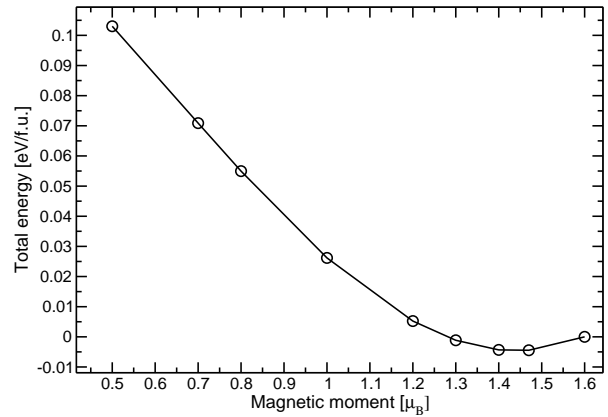}
\caption{\label{FSM}   
The total energy dependence on the spin moment, obtained in the
fixed spin moment calculations in the LSDA+U$_{Co}$ approximation.
}
\end{figure}

In order to check that the solution obtained corresponds to the global
minimum of the density functional in the LSDA+U approximation the
fixed spin moment calculations were performed. One may see
in Fig.~\ref{FSM} that the total energy of the system drastically
grows with decrease of the spin moment making the state with S=1/2
(low spin state of the Ru$^{5+}$ ions), proposed in Ref.~\onlinecite{Zhou2012}, 
energetically unfavorable. However the minimum $E(\mu)$ by itself is 
flat enough, implying that the spin fluctuations may be 
operative in Ba$_3$CoRu$_2$O$_9$.

The intradimer exchange coupling was found to be antiferromagnetic,
$J_{intra}=211$~K, for the Heisenberg model as defined in Eq.~\ref{Heisenberg}
with $S=3/2$. Each Ru--Ru dimer is connected with three other dimers on each 
side (i.e. with six dimers in a sum) via CoO$_6$ octahedra. The exchange coupling between 
the nearest Co and Ru ions ($J_{Co}$) is small, and does not exceed 8~K. The
coupling between dimers is larger $J_{inter}$: 30.4, 30.4 and 16.6~K.
So that in the LSDA+U$_{Co}$ Ba$_3$CoRu$_2$O$_9$ should be considered 
as a system of coupled dimers. 

On the second step we added the $U-$correction for the Ru $4d$ states
with U = 3 eV and J$_H=0.7$ eV, so that both Ru $4d$ and Co $3d$
states were considered as correlated (abbreviated as LSDA+U$_{Co,Ru}$). 
As a result both the  magnetic moment on Ru and the band gap grew in absolute 
value. The spin moments were found to be 
$|m_{Ru}| =1.95$~$\mu_B$, $|m_{Co}| =2.61$~$\mu_B$,
while the band gap equals 1.11 eV. The exchange constants are the following:
$J_{intra}=150$~K,  $J_{inter,1}=39.8$~K, $J_{inter,2}=39.8$~K, 
$J_{inter,3}=19.4$~K.

Comparing calculated and experimental values of the local magnetic moments, one 
may see that while the LSDA+U significantly improves
the calculation results when the $U-$correction is applied only to the 
Co 3$d$ states, but makes an agreement with experiment worse, if it 
is used both for the Ru $4d$ and Co $3d$ states.

The LSDA+U method was designed to describe electronic and magnetic
properties of the 3$d$ transition metal compounds. In the traditional 
realization of this method for description of the electronic properties of the 
transition metal oxides the $U-$correction is applied only to the $d$
wave functions: 
\begin{equation}
H_{LSDA+U} = H_{LSDA} + \sum_{mm'} | \psi_{inlm\sigma}\rangle 
V^{\sigma}_{mm'} \langle \psi_{inlm'\sigma} |.
\end{equation}
Here $i$ is the site index, $n,l,m$ are the principal, orbital
and magnetic quantum numbers, $H_{LSDA}$ is the LSDA Hamiltonian, and 
$V^{\sigma}_{mm'}$ is the $U-$correction as defined in Ref.~\onlinecite{Anisimov1997}.
In the LMTO method $\psi_{inlm\sigma}$ are the corresponding linearized muffin-tin (LMT) 
orbitals~\cite{Anisimov1991a,Andersen1984} for $d$ states, in the linearized 
augmented plane wave (LAPW) method -- the ``muffin-tin'' part of a
radial function for the $d$ orbital.~\cite{shick1999}

However, in the compounds based on the $4d$ and $5d$ transition metal
ions or even $3d$ ions with a high oxidation state the electrons of interest
are localized on the orbitals which significantly differ from the atomic $d$ wave 
functions (see e.g. Fig.~6 in Ref.~\onlinecite{Ushakov2011}, where the orbital 
on which one of the Au $5d$ electrons localizes in Cs$_2$Au$_2$Cl$_6$ is shown). 
An application of the Hubbard's like $U-$correction only to
the $d$ part of the wave function is methodologically incorrect. The use of the Wannier 
functions centered on the transition metal ions, but with large contributions on the 
surrounding atoms, is more appropriate in this case. Corresponding
version of the LDA+U method in the Wannier functions basis set was
recently proposed.~\cite{Korotin2012}

The squared coefficients of the Wannier function expansion in terms of 
the LMT-orbitals show the contribution of each orbital.~\cite{Anisimov-05} 
In the case of Ba$_3$CoRu$_2$O$_9$ the Wannier functions, centered on the Ru ions and
corresponding to the LDA bands expanded from -0.5 eV to 0.8 eV, have 
the contribution coming from the Ru $4d$ LMT-orbitals of 55$\%$ only,
while $\sim$40 $\%$ corresponds to the O $2p$ and $\sim$ 5 $\%$ to the Co $3d$
orbitals. Thus, applying the $U-$correction on the Ru 4$d$ states we
acting only on the part of the wave function and force the electrons
to localize on the atomic $d$ orbitals, not on the Wannier functions preferred by
the LDA. This results in the overestimation of the spin moment on Ru and the band 
gap in the LSDA+U$_{Co,Ru}$ calculations. Nonphysical increase of the 
$U-$parameter on Ru up to 12 eV leads to the Ru spin moment 2.74 $\mu_B$ close to 
what one would expect for the $d^3$ configuration, if electrons localize on the 
atomic $d$ orbitals.

Thus, we see that a small value of the spin moment on Ru, $1.47$ $\mu_B$,
in the LSDA+U$_{Co}$ calculation is related to the fact that the
electrons are indeed localized on the Wannier functions, which have 
substantial contribution from the O 2$p$ states. A naive estimation of the Ru spin moment 
from the Ru $4d$ contribution to the LDA Wannier functions $3\times 0.55 = 1.65$ is close 
to what we obtain in the real LSDA+U$_{Co}$ calculation.

An alternative mechanism of the magnetic moment reduction on Ru is a stabilization
of the ``orbital selective spin singlet'' state. The large bonding-antibonding 
splitting may lead to the formation of the spin singlets on the $a_{1g}$ orbitals,
while the electrons occupying $e_g^{\pi}$ orbitals may stay localized and bear
magnetic moment (2 $\mu_B$), which can be again reduced by the hybridization with
oxygen. The intra-atomic Hund's rule exchange will act against this
state, but a final answer should be given by a direct calculation. The one-electron 
approximations like the LSDA or LSDA+U are useless in this situation, since
they are not able to simulate the spin singlet state, which should
be described by a true many-particle wave function. The cluster 
dynamical mean field theory (DMFT) should be used instead.

\section{Conclusions}
With the use of the LSDA+U calculations (with the $U-$correction applied on the 
Co $3d$ states) we show that the electronic ground state in the low temperature 
phase of Ba$_3$CoRu$_2$O$_9$  follows the Hund's rule with
$(t_{2g} \uparrow)^3$ electronic configuration (all Ru $d$ electrons on one site 
have the same spin). This is contrast to the unusual orbital filling 
$(t_{2g} \uparrow)^2 (t_{2g} \downarrow)^1$ proposed previously basing
on the crystal structure analysis.~\cite{Zhou2012} Very similar situation
was observed in the case of La$_4$Ru$_2$O$_{10}$, where the stabilization
of the low spin state of the Ru$^{4+}$ was first proposed experimentally.~\cite{Khalifah2002}
However following band structure calculations together with the X-ray
measurements showed that this is unlikely.~\cite{Wu2006}  
In  Ba$_3$CoRu$_2$O$_9$ the suppression of the magnetic moment on the Ru ions from 
3~$\mu_B$ expected from the naive
atomic consideration to 1.17-1.45~$\mu_B$ observed in the experiment~\cite{Rijssenbeek1999,Lightfoot1990} 
is attributed the strong hybridization effects. Due to a large spatial 
expansion of the $4d$ wave functions and a high oxidation state of Ru (5+)
the Ru $4d$ and O $2p$ states are strongly hybridized. As a result three $d$
electrons of Ru$^{5+}$ localize not on the atomic, but on the Wannier 
orbitals with significant contributions from the spin non-polarized 
O $2p$ states. We also show that the LSDA+U approximation must be applied
with care for the description of the $4d$ and $5d$ transition metal compounds.

\section{Acknowledgments}
I am grateful to Prof. D. Khomskii for the various discussions 
about the suppression of the magnetic moment in the dimer and trimer systems.
The calculations were performed on the ``Uran'' cluster.
This work is supported by the Russian Foundation for Basic Research 
via grant RFFI-13-02-00374 and by the Ministry of 
education and science of Russia  through the programs 12.740.11.0026 and MK-3443.2013.2.

\bibliography{../library}
\end{document}